\documentclass[%
superscriptaddress,%
twocolumn,%
prl,%
aps,%
showpacs,%
preprintnumbers,%
amsmath,%
amssymb,%
twoside,%
floatfix,
]{revtex4-1}

\usepackage{graphicx}       
\usepackage{xspace}

\newcommand{\ttas}{1$T$-TaS$_\mathrm{2}$\xspace}
\newcommand{\tbt}{$\sqrt{13} \times \sqrt{13}$\xspace}

\newcommand{\ts}{$\mathbf{T}_S$\xspace}
\newcommand{\ef}{E$_\mathrm{F}$\xspace}

\begin{document}


\title{Orbital textures and charge density waves in transition metal dichalcogenides}


\author{T.~Ritschel} \affiliation{Leibniz Institute for Solid State and
Materials Research IFW Dresden, Helmholtzstr. 20, 01069 Dresden, Germany}
\affiliation{Institute for Solid State Physics, Dresden Technical University,
TU-Dresden, 01062 Dresden, Germany}

\author{J.~Trinckauf} \author{K.~Koepernik} \affiliation{Leibniz Institute
for Solid State and Materials Research IFW Dresden, Helmholtzstr. 20, 01069
Dresden, Germany}

\author{B.~B\"uchner} \affiliation{Leibniz Institute for Solid State and
Materials Research IFW Dresden, Helmholtzstr. 20, 01069 Dresden, Germany}
\affiliation{Institute for Solid State Physics, Dresden Technical University,
TU-Dresden, 01062 Dresden, Germany}

\author{M.~v.~Zimmermann} \affiliation{Deutsches Elektronensynchrotron DESY,
Notkestr. 85, 22603 Hamburg, Germany}

\author{H.~Berger} \affiliation{Ecole polytechnique Federale de Lausanne,
Switzerland}

\author{Y. I. Joe} \author{P. Abbamonte} \affiliation{Department of Physics and
Frederick Seitz Materials Research Laboratory, University of Illinois, Urbana,
Illinois 61801, USA }

\author{J.~Geck} \affiliation{Leibniz Institute for Solid State and Materials
Research IFW Dresden, Helmholtzstr. 20, 01069 Dresden, Germany}


\begin{abstract}
  Low-dimensional electron systems, as realized in layered materials, often tend
  to spontaneously break the symmetry of the underlying nuclear lattice by
  forming so-called density waves~\cite{Gruner1994a}; a state of matter that
  currently attracts enormous
  attention~\cite{Chang2012,Ghiringhelli2012,SilvaNeto2014,Cruz2008,Muehlbauer2009}. 
  Here we reveal a remarkable and surprising feature of charge density waves
  (CDWs), namely their intimate relation to orbital order. For the prototypical
  material \ttas we not only show that the CDW within the two-dimensional
  TaS$_2$-layers  involves previously unidentified orbital textures of great
  complexity. We also demonstrate that two metastable stackings of the orbitally
  ordered layers allow manipulation of salient features of the electronic structure.
  Indeed, these orbital effects provide a route to switch \ttas nanostructures from
  metallic to semiconducting with technologically pertinent gaps of the order of
  200\,meV. This new type of orbitronics is especially relevant for the ongoing
  development of novel, miniaturized and ultra-fast devices based on layered
  transition metal dichalcogenides~\cite{Wang2012,Chhowalla2013}.
\end{abstract}

\maketitle


\begin{figure}[]
  \centering
  \includegraphics[width=\columnwidth]{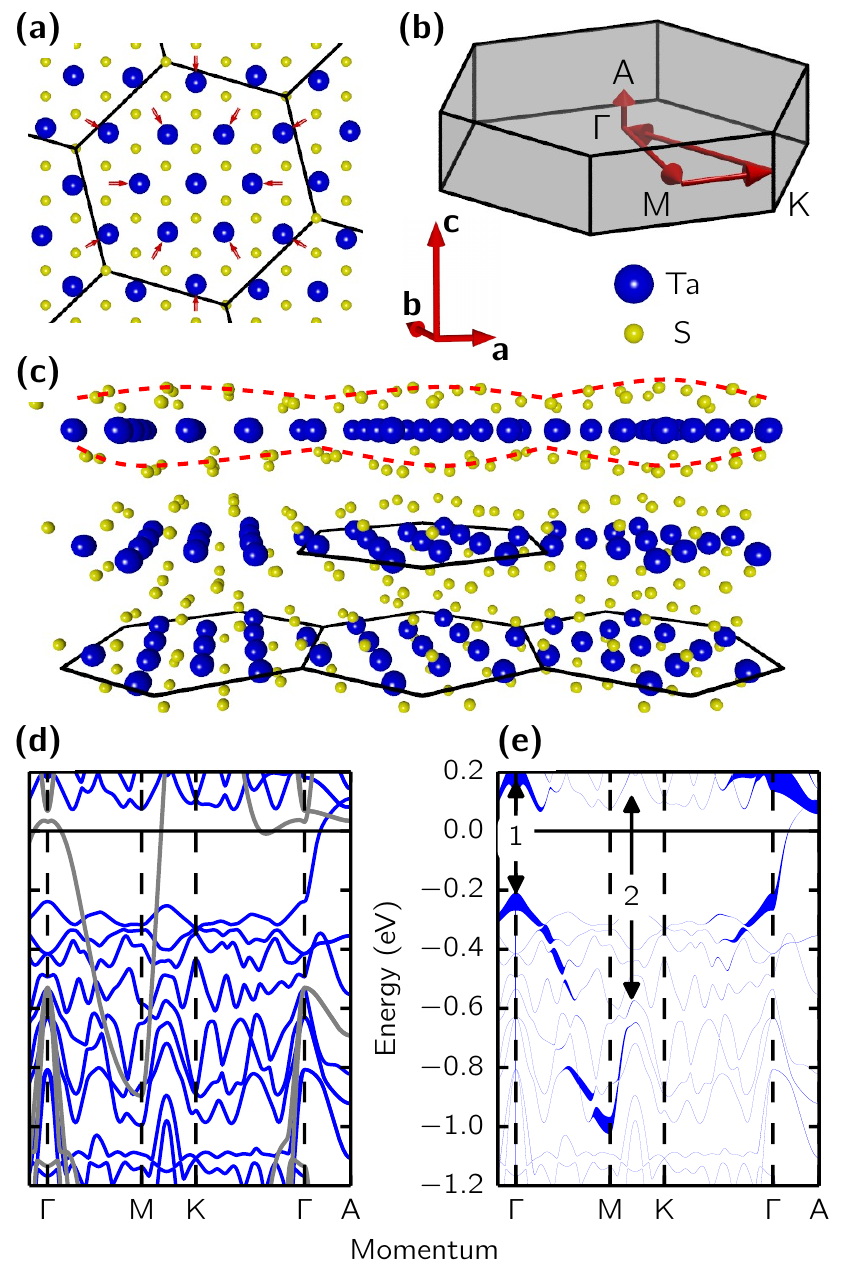}
  \caption{\textbf{The \tbt supercell structure of \ttas.}
    (a) and (c): View along the $c$-axis and parallel to the $ab$-planes,
    respectively ($\mathbf{a},\mathbf{b}$,$\mathbf{c}$:  lattice vectors of the
    undistorted  P$\bar3$m1 crystal structure).
    The Ta-displacements indicated by red arrows in (a) are mainly  parallel to
    the $ab$-planes, resulting in clusters containing 13 Ta-sites.  Red dashed
    lines in (c) highlight the breathing of the S-sites perpendicular to the
    $ab$-plane.  (d): LDA supercell band structure along the high symmetry
    direction in the Brillouin zone shown in (b).  The gray lines indicate the
    unreconstructed band structure. (e): Unfolded band structure.  The
    thickness of bands measures the spectral weight at this position.  Arrows
    indicate the gaps at $\Gamma$ and between M and K which are commonly named
    Mott-gap (1) and CDW-gap (2), respectively.} 
\end{figure}
Among the various transition metal dichalcogenides (TMDs), \ttas stands out because of its particularly rich electronic phase diagram as a function of pressure and temperature~\cite{Sipos2008}. This phase diagram not only features incommensurate, nearly commensurate and commensurate CDWs, but also pressure-induced superconductivity below 5~Kelvin. 
In addition to this, it was proposed early on that the low-temperature commensurate CDW (C-CDW), which is illustrated in Fig.~1~(a),(c), also features many-body Mott-physics~\cite{Fazekas1979}.
Experimental evidence for the presence of Mott-physics in \ttas has indeed been obtained recently by time-resolved spectroscopies, which observed the ultra-fast collapse  of a charge excitation gap, which has been interpreted as a fingerprint of significant electron-electron interactions~\cite{Hellmann2010a,Petersen2011,Perfetti2006}. 

Even though the above scenario for the C-CDW is widely accepted, important experimental facts remain to be understood: the very strong suppression of the C-CDW with external pressure is puzzling. Already above 0.6~GPa, the C-CDW is no longer stable, although nesting conditions, band widths as well as the lattice structure remain essentially unchanged.  It is also not clear how ordered defects within the C-CDW order, which emerge in the nearly commensurate phase (NC-CDW) upon heating~\cite{Wu1989,Spijkerman1997} and which do not cause significant changes in the bandwidths, can render the electron-electron onsite interaction U completely ineffective~\cite{Ang2012}. 
In the following we will show that all these issues are explained consistently in terms of orbital textures that are intertwined with the CDW. In addition to this, we also demonstrate that this new twist to the physics of CDWs also provides a new and powerful device concept for future applications based on TMDs.


\begin{figure*}[]
  \centering
  \includegraphics[width=.96\textwidth]{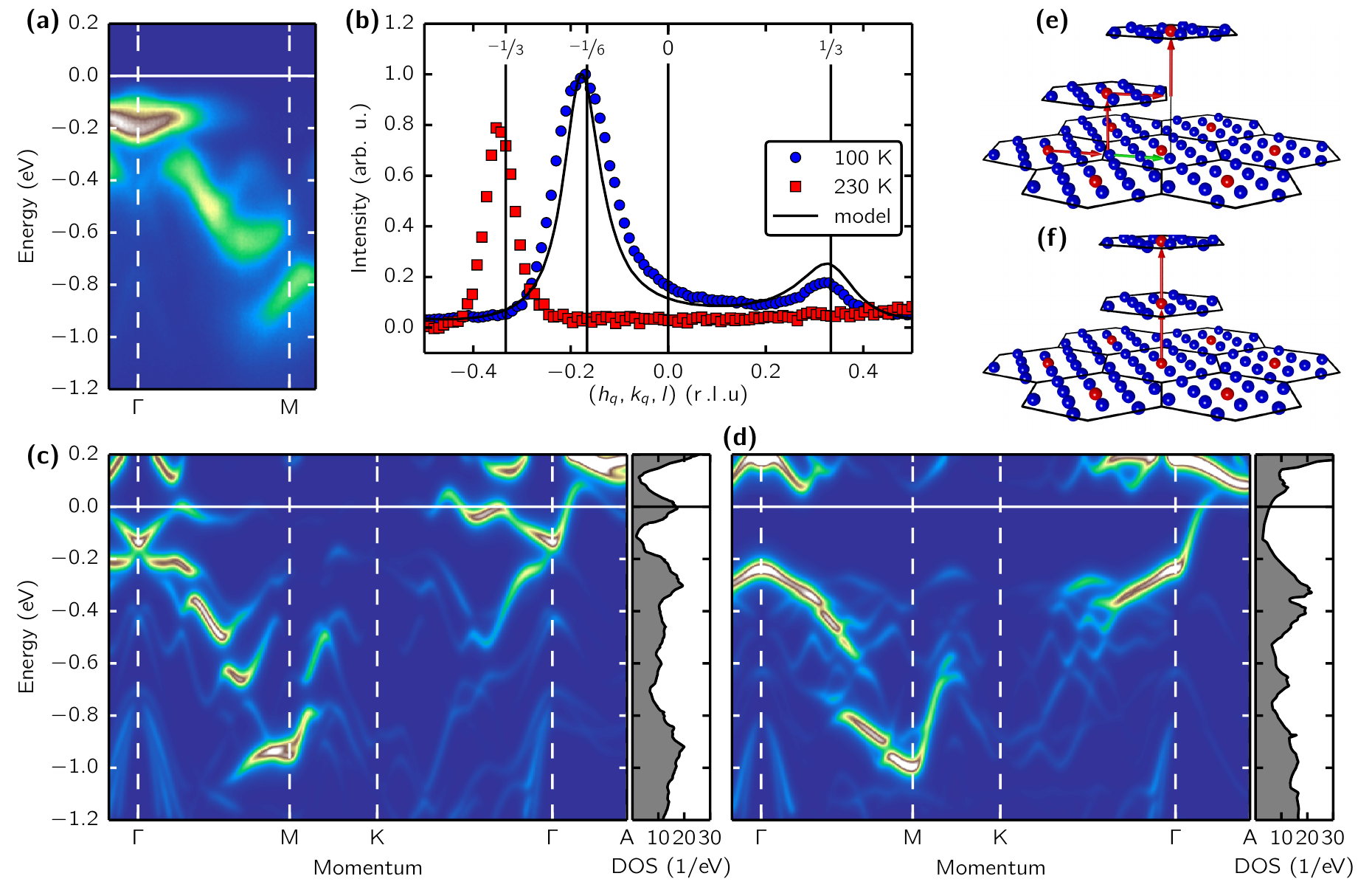}
  \caption{\textbf{Different layer stackings and their impact on the band
    structure.} (a): ARPES data of \ttas measured at 1\,K. (b): XRD intensity of
    \ttas measured along the $l$-direction for the C-CDW and NC-CDW phase. The
    in-plane $q$-vector corresponds to $h_q = 16/13$, $k q = 1/13$ for the C-CDW
    phase at
    100\,K and $h q = 1.248$, $k q = 0.068$ for the NC-CDW phase at 230 K. The single
    reflection observed for the NC-CDW implies a well-ordered stacking along $c$,
    fully consistent with previous results~\cite{Spijkerman1997}. On entering the C-CDW phase, the XRD
    pattern changes markedly and two broad peaks appear owing to the presence of an
    alternating $c$-axis stacking (see text). A numerical simulation of the XRD (solid
    line) further reveals a finite correlation length of the alternating $c$-axis
    stacking of approximately $10 \times c$. (c) and (d): Calculated band
    structure for $\mathbf T_S = 2\mathbf{a}+\mathbf{c}$ and $\mathbf T_S
    = 1\mathbf{c}$, respectively. 
    To facilitate the comparison with ARPES data shown in (a) the bandweights
    have been convolved with a Lorentzian shaped resolution function. (e) and
    (f): Visualization of the stacking with $\mathbf T_S
    = 2\mathbf{a}+\mathbf{c}$  and $\mathbf T_S = 1\mathbf{c}$, respectively.
    Red arrows indicate the stacking vectors that connect the central Ta-sites
    (shown in red) in successive layers.
  }
\end{figure*}

An important feature of the C-CDW is its partially disordered stacking along the $c$-axis:
characterizing the relative alignment of the C-CDW in adjacent $ab$-planes in terms of stacking vectors \ts, which connect the central Ta-sites in successive layers (cf. Figs.~2~(e) and (f)), 
the $c$-axis stacking is given by an alternation of $\mathbf T_S=1\mathbf{c}$ with a random choice from the three symmetry equivalent vectors $\mathbf T_S=2\mathbf a + \mathbf c$, $2\mathbf b +\mathbf c$ or $-2(\mathbf a + \mathbf b)+\mathbf c$~\cite{Tanda1984,Nakanishi1984}.
The effect of the different \ts on the electronic structure so far remained unexplored. Our highly efficient DFT approach (see methods), however, enables us to study these effects for the first time within the local density approximation (LDA).

For our {\itshape ab initio} calculations, we chose two superstructures with $\mathbf{T}_S=1\mathbf{c}$ and $\mathbf{T}_S = 2\mathbf a + \mathbf c$, respectively, which represent the two metastable types of stacking found experimentally. 
The calculated total energy for both superstructures differs only by a few meV, implying that these two stacking types are indeed metastable.
Fig.~1~(d) shows the band structures for the C-CDW with $\mathbf{T}_S=1\mathbf{c}$ and the undistorted lattice, where the band dispersions are shown along high symmetry directions of the Brillouin zone of the unmodulated structure (Fig.~1~(b)).
In good agreement with previous DFT results~\cite{Bovet2003}, the band structure in the plane through $\Gamma$, M and K is fully gapped and only along $\Gamma$-A a Fermi level crossing occurs. This yields a one-dimensional metal, in line with recent reports~\cite{Darancet2014}.
A major disadvantage of such supercell calculations is, however, that the additional backfolded bands only indicate for which momenta $\mathbf{k}$ and energies $\omega$ electronic states exist. There is no information about the spectral weight of these states. Exactly this information is provided by the so-called spectral function $A(\mathbf{k},\omega)$~\cite{Damascelli2003}, which is the physically relevant quantity that can be accessed experimentally by~ e.g., angle-resolved photoemission spectroscopy (ARPES).

An approximation for  $A(\mathbf{k},\omega)$ within DFT, can be obtained using the so-called unfolding procedure~\cite{Ku2010}. 
The result of this unfolding procedure for the band structure shown in Fig.~1~(d) is presented in Fig.~1~(e), where the thickness of the individual bands reflects the spectral weight $|A(\mathbf{k},\omega)|^2$. It is clearly visible that only a small fraction of the reconstructed band structure carries substantial spectral weight.
Unlike the bands in Fig.~1~(d), the unfolded band structure allows to clearly identify dispersing bands and, correspondingly,  the size and location of energy gaps. Specifically, our calculation yields two gaps around the Fermi level \ef, indicated by arrows in Fig.~1~(e). The gap at $\Gamma$ is usually interpreted as a Mott-gap due to the electron-electron interaction U~\cite{Perfetti2006,Rossnagel2011}, while the gap between M and K fits well to the Fermi surface nesting vector and is therefore commonly assigned to a CDW-gap due to the electron-phonon coupling. 
It is very important to note here, that our calculations do not include any onsite interactions U. The fact that the gap at $\Gamma$ is obtained without including U is therefore surprising and already implies that this gap cannot be caused by Mott-physics alone.

In  Fig.~2 we compare 
the unfolded LDA band structures for $\mathbf{T}_S=1\mathbf{c}$ and $\mathbf T_S = 2\mathbf a +\mathbf c $, which yields the first major result of this study: the dramatic dependence of the electronic structure on the CDW-stacking. 
As can be observed in Figs.~2~(c) and (d), essential features of the low-energy electronic structure depend critically on the stacking. Most notably the bands around the $\Gamma$-point are strongly affected. While for $\mathbf{T}_S = 2\mathbf{a}+\mathbf{c}$ the in-plane dispersions show \ef-crossings along $\Gamma$-K and $\Gamma$-M, both crossings are completely absent for $\mathbf{T}_S=1\mathbf{c}$. 
%
%
Therefore, a transition from an in-plane metal to an in-plane semiconductor occurs as a function of \ts. Considering that the $ab$-planes of \ttas are usually thought to realize strongly two-dimensional metallic systems, these are most surprising results. 

The comparison of the two DFT models to the ARPES data in Fig.~2~(a) shows that the calculation for $\mathbf{T}_S = 2\mathbf{a}+\mathbf{c}$ describes the experiment very well for binding energies below -0.3~eV.  
However, it is also obvious that both DFT models fail to describe the electronic states close to \ef and, in particular, the pseudo-gap observed experimentally~\cite{Pillo1999}. Including an onsite U at Ta does not cure this discrepancy, implying that these effects are beyond LDA+U.
Indeed, the strong \ts-dependence of the electronic structure suggests that these deviations are related to the stacking disorder in the real material~\cite{Nakanishi1984}, which cannot be taken into account by our DFT-models.
While this issue certainly deserves scrutiny in future work, here we focus on determining the origin of the dramatic effects of $\mathbf T_S$.

To this end we calculated the charge density distribution of the uppermost occupied states in real space. As can be seen in Fig.~3, for both stackings a complex orbital texture within the $ab$-plane emerges. 
This is the second major result of this work: the discovery of an orbital texture that is intertwined with a CDW. Note that not only the occupancy of a certain type of orbital changes spatially. Instead also the symmetry of the orbitals clearly changes from site to site, resulting in a complex orbital ordering pattern. 

The orbital structure for $\mathbf{T}_S=1\mathbf{c}$ permits significant charge hopping only along $c$, as illustrated in Fig.~3~(e). In other words, the charges flow along orbital stripes along $c$, corresponding to the quasi one-dimensional character of the uppermost band in Fig.~3~(a) and Fig.~2~(d).
This drastically changes in the case of $\mathbf{T}_S = 2\mathbf{a}+\mathbf{c}$. Now there is a significant $ab$-component of the hopping as illustrated in Fig.~3(f), so that the uppermost band attains a larger in-plane dispersion and crosses \ef along $\Gamma$-M and $\Gamma$-K (Figs.~3~(b) and 2~(c)).

\begin{figure}
  \centering
  \includegraphics[width=\columnwidth]{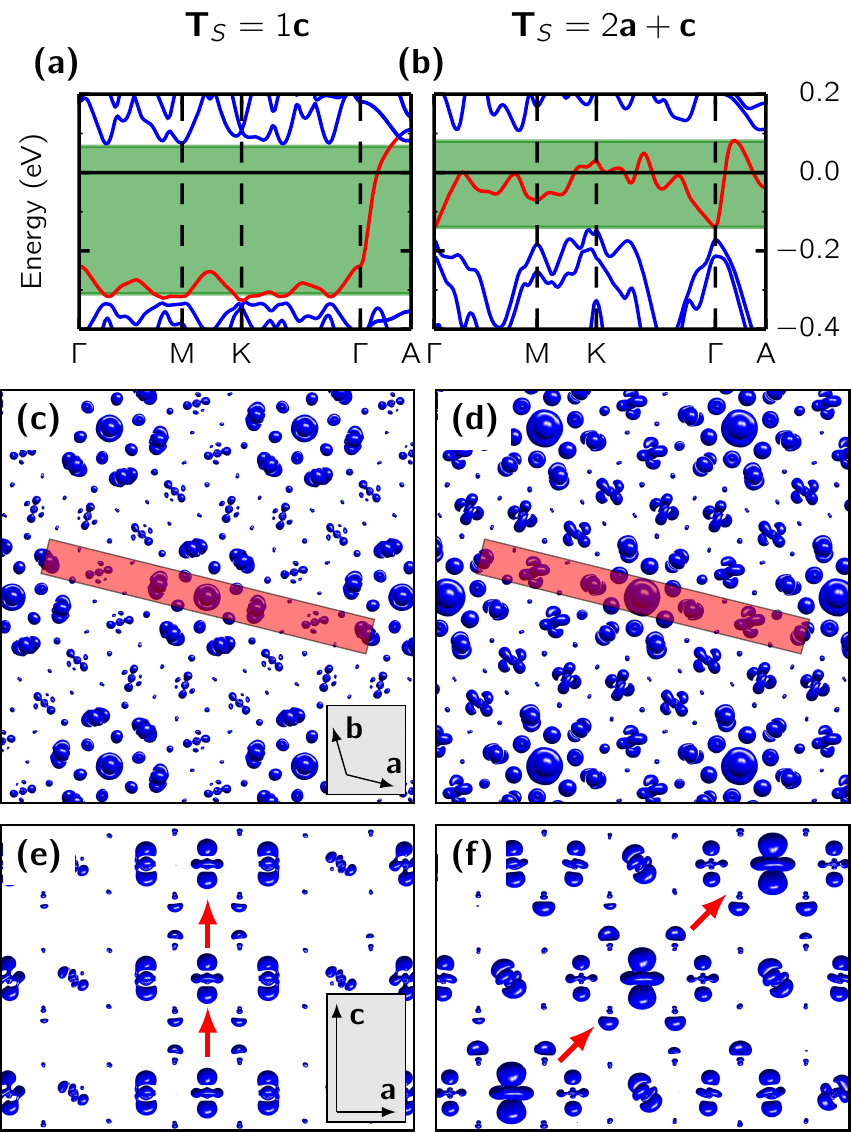}
  \caption{\textbf{Real space illustration of the electron density for the
    highest occupied band.} (a) and (b): The band structure near \ef. The energy
    window used to calculate the energy-resolved electron density is indicated
    by the green area.  (c) and (d): A complex orbital texture emerges within
    the $ab$-plane for $\mathbf T_S = 1\mathbf c$ (c) and $\mathbf T_S
    = 2\mathbf a + \mathbf c$. (e) and (f), View of the $ac$-plane corresponding
    to the red areas in (c) and (d). (e): For $\mathbf T_S = 1\mathbf c$
    significant hopping is allowed only along the $c$-direction.  (f):
    A substantial $ab$-component of the hopping appears for $\mathbf T_S
    = 2\mathbf a + \mathbf c$.
  }
\end{figure}
These results have important consequences: Firstly, the comparison of the ARPES data and the DFT results shows that the $\Gamma$-gap is obtained on a quantitative level without an onsite repulsion U on Ta. Although this result does not exclude the presence of electron-electron interactions, it strongly argues against U being the main cause for the $\Gamma$-gap. Instead, the band structure calculations reveal that this gap is for a large part due to the interlayer hybridization.
The latter is in turn mostly caused by the  $3z^2-r^2$-type orbitals pointing along $c$, i.e., the $\Gamma$-gap is directly coupled to the orbital texture. 
This result naturally explains the ultra-fast response of this gap observed in time-resolved ARPES experiments, since the disruption and reordering of the electronic orbitals can evolve on much faster time-scales than the lattice. 
Secondly, the presence of orbital textures explains  the strong pressure dependence of the CDW-order in \ttas, because through the orbital texture pressure has a large effect on the stability of CDW-phases with different stackings in a way that goes well beyond traditional nesting scenarios. 
We also stress that the crucial role of the interlayer interactions is verified experimentally by the data in Fig.~2~(b), which shows that the stacking changes completely across the C-CDW/NC-CDW transition. This change in stacking together with our DFT-results also rationalizes the collapse of the gap at $\Gamma$ upon entering the NC-CDW with warming.

\begin{figure}[t!]
  \centering
  \includegraphics[width=\columnwidth]{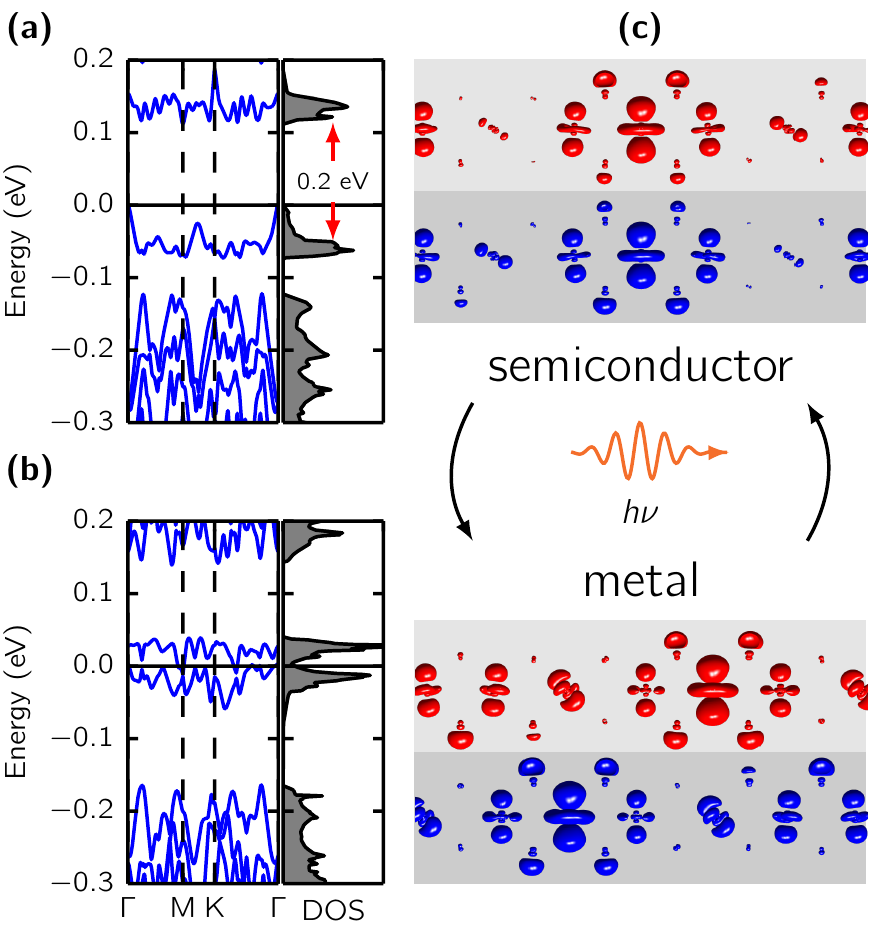}
  \caption{
    \textbf{Device concept based on the switching between metastable orbital orders.}
    Calculated band structure for a bilayer of \ttas with $\mathbf T_S
    = 1\mathbf c$ (a) and $\mathbf T_S = 2\mathbf a + \mathbf c$ (b). 
    A semiconductor-to-metal transition takes place upon changing the stacking of the two layers.
    (c): Orbital order corresponding to the semiconducting (top) and metallic state (bottom). 
  }
\end{figure}

Thirdly and most importantly, the relative orientation of the orbitals in adjacent $ab$-planes has a spectacular effect on the band dispersions. This can be readily understood in terms of  the overlap integrals, which depend critically on the relative orientation of the orbitals in adjacent layers along $c$ (cf.~Figs.~2~(e),(f)).
The above immediately yields a new device concept, which employs metastable orbital orders for controlling the electronic structure of nanostructures:
As illustrated in Fig.~4 for a bilayer of \ttas, switching between the metastable orbital configurations causes a complete semiconductor-metal transition. In other words, by changing the orbital order in the direction \textit{perpendicular} to the layers, one can control the conductivity \textit{parallel} to the layers.
Even though this effect has not yet been observed directly, a very recent
experiment provides first evidence that this might be achieved reversibly and on
ultra-fast timescales using optical laser pulses~\cite{Stojchevska2014}.

Orbital textures hence enable to manipulate the band dispersion and gap structure of \ttas in a very effective way. They therefore provide a new route to tailor and switch the electronic properties of TMDs, possibly on the femtosecond time scale. This concept of orbitronics may hence enable to create novel, small and ultra-fast electronics, which adds to the large potential of TMDs for future device applications.


\section{Methods}
The \textbf{DFT} calculations were done using the FPLO14
package~\cite{Koepernik1999}, which has been developed at the IFW Dresden and supports unfolding of the band
structure as outlined in Ref.~\onlinecite{Ku2010}. Due to its small basis size it allows for calculations of large supercells extremely efficiently, which makes the presented calculations numerically effordable. 
The supercell structure for
$\mathbf T_S = 1\mathbf c$ was derived from Ref.~\onlinecite{Brouwer1978}~and~\onlinecite{Spijkerman1997}.  
In order to simulate different layer stackings we
transformed the hexagonal supercell into a triclinic supercell without changing
the atomic displacements. It is important to note that a structural relaxation, which starts from the undistorted lattice, rapidly converges against the experimentally observed superstructure. 
In addition, the present DFT calculations reproduce the so-called Mott and CDW gap on a quantitative level.
This verifies our approach and shows that the present DFT-models capture the important interactions present in the real material.

\textbf{XRD} data were obtained at the beamline BW5 at the Deutsches
Elektronensynchrotron (DESY).  We mounted a high-quality single crystal
of \ttas in a displex Helium cryostat sitting in an Euler cradle.
The presented XRD data was obtained by measuring the diffracted intensity of the
superstructure reflection while scanning along the crystallographic
$l$-direction for two different temperatures.

\textbf{ARPES} measurements were conducted at the $1^3$-ARPES endstation at beamline
UE112PG2 at the Berlin Synchrotron (BESSY).  We used $p$-polarized light of
96~eV photon energy, so that the final state crystal momentum at normal
emission corresponds to the $\Gamma$-point~\cite{Rossnagel2005}.  The sample
temperature was kept at 1~K.



\section{Acknowledgments}

This work was financially supported by the German Research Foundation under
grant DFG-GRK1621.  J.T and J.G gratefully acknowledge the financial support by
the German Research Foundation through the Emmy Noether program (grant
GE~1647/2-1). 
Y.I.J. and P.A. were supported by U.S. Department of Energy grant DE-FG02-06ER46285.
We thank K.~Rossnagel for fruitful discussions.




\end{document}